# Ground and Excited Exciton Polarons in Monolayer MoSe$_2$

Thomas Goldstein[1], Yueh-Chun Wu[1], Shao-Yu Chen[1,2], Takashi Taniguchi[3], Kenji Watanabe[4], Kalman Varga[5] and Jun Yan[1*]

[1]Department of Physics, University of Massachusetts, Amherst, Massachusetts 01003, USA

[2]School of Physics and Astronomy, Monash University, Clayton Campus VIC 3800, Australia

[3]International Center for Materials Nanoarchitectonics,
National Institute for Materials Science, Tsukuba, Ibaraki 305-0044, Japan

[4]Research Center for Functional Materials,
National Institute for Materials Science, Tsukuba, Ibaraki 305-0044, Japan

[5]Department of Physics and Astronomy, Vanderbilt University, Nashville, Tennessee 37235, USA

*Corresponding Author: Jun Yan. E-mail: yan@physics.umass.edu

Tel: (413)545-0853 Fax: (413)545-1691

**Abstract:**

Monolayer transition metal dichalcogenide semiconductors, with versatile experimentally accessible exciton species, offer an interesting platform for investigating the interaction between excitons and a Fermi sea of charges. Using hexagonal boron nitride encapsulated monolayer MoSe$_2$, we study the impact of charge density tuning on the ground and excited Rydberg states in the atomic layer. Consistent exciton-polaron behavior is revealed in both photoluminescence and reflection spectra of the A exciton 1s (A:1s) Rydberg state, in contrast to previous studies. The A:2s Rydberg state provides an opportunity to understand such interactions with greatly reduced exciton binding energy. We found that the impact of the Fermi sea becomes much more dramatic. With a photoluminescence upconversion technique, we further verify the 2s polaron-like behavior for the repulsive branch of B:2s exciton whose energy is well above the bare bandgap. Our studies show that the polaron-like interaction features are quite generic and highly robust, offering key insights into the dressed manybody state in a Fermi sea.

The atomically thin crystals of hexagonal transition metal dichalcogenide (H-TMD) semiconductors have attracted considerable interest in recent years due to their unique optical, electronic, spintronic and valleytronic properties [1,2]. These materials, composed of covalently bonded layers of chemical form $MX_2$ stacked weakly atop each other, become non-centrosymmetric direct-gap semiconductors with band edges located at the +/-K corner points (valleys) of the Brillouin zone when reduced to a single layer (1L) in thickness [3,4]. The innate broken inversion symmetry in H-TMD monolayers, together with the spin-orbit interaction, lifts the spin degeneracy at +/-K valleys, giving rise to two species of spin-zero bright excitons A and B, involving electrons and holes of opposite spins in each valley (filled and open circles in Fig.1a) [3,4]. Isolated from the bulk material and placed on a substrate with smaller dielectric constant, such as quartz, $SiO_2$ or hexagonal boron nitride (hBN), 1L-TMDs manifest enhanced Coulomb interaction effects. The A and B excitons have large binding energies of a few hundred meV, opening up a large energy window to explore not only ground, but also excited Rydberg exciton states [5–14].

The objective of this Letter is to study the interaction effects of the A and B Rydberg excitons with a Fermi sea of charges. As a two-dimensional (2D) materials system, the charge density in 1L-TMD semiconductors can be continuously tuned with an external gate. The impact of charge doping on the A exciton 1s (A:1s) ground state, the most prominent optical feature in 1L-TMD, has been extensively studied and widely interpreted with the formation of trions [15–19]. Trions are composite particles composed of either two electrons and one hole, or two holes and one electron, and they form only when the sample is doped. Meanwhile in a doped sample, a subtle question regarding the interaction between an exciton and a Fermi sea of electrons or holes, i.e. the exciton-polaron problem, versus just one electron or one hole, arises[20,21]. Early studies suggest that the trion picture can be used to explain TMD photoluminescence (PL) emission for devices with Fermi energy $E_F$ up to tens of meV, where the effect of Fermi sea can be accounted by Pauli blocking [15]. Later reflection spectroscopy studies of doped 1L-TMDs attributed A:1s absorption features as exciton-polarons [22–24]. In some of these devices PL was also measured, but due to their large Stokes shift from absorption, these PL features were believed not resulting from exciton-polarons [22].

The interactions between excited Rydberg excitons and charges are more elusive. While neutral excited Rydberg excitons have been measured using a variety of techniques[5–14], exploration of their interaction with a Fermi sea of charges (Fig.1b) is so far limited. [19,25–28] As excited states, these excitons have more decay channels and thus shorter lifetimes. Due to their smaller binding energy, in combination with renormalization effects in the presence of a Fermi sea, the spectral function of such excited states is believed to merge with the free electron-hole continuum rapidly[20,21]. Excited trion state was theoretically calculated by numerically solving exciton-electron Schrödinger equation, where it was found that the existence of 2s trions requires an electon-hole mass ratio larger than 16, not realized in TMDs.[25] A recent experimental and theoretical work on $WS_2$, on the other hand, reported signatures of 2s trions in absorption spectroscopy.[19] Another theoretical calculation found evidence of 2p trions instead of 2s trions.[26]

Here using PL and differential reflection (DR) spectroscopy, we investigate exciton-polarons in monolayer molybdenum diselenide ($MoSe_2$). The high-quality sample produces a panoply of optical features due to A:1s, A:2s, B:1s and B:2s states, allowing for a comprehensive examination of interaction effects arising from different exciton species coupling to a Fermi sea of holes. In contrast to previous studies[22–24], we find that PL, with minimal Stokes shift from DR, provides a very powerful tool to study the polaronic dressing effects. We show that the energy and linewidth evolution of A:1s PL provide a comprehensive test of the exciton-polaron theory[20,21]. With A:2s PL, we examine how the decrease of the exciton binding energy impacts the exciton polarons. Quite remarkably, despite large Landau damping, a hint of polaronic effect is seen even for B:2s exciton whose energy is well above the bare bandgap. Our results provide extensive evidence that qualitative features of polaron interaction are highly robust, and that quantitatively, the dressing effect is very sensitive to the exciton state involved.

Figure 1c shows an optical microscope image of our sample. We grow bulk $MoSe_2$ crystals via the chemical vapor transport method using chlorine as a transport agent.[29,30] The monolayer sample was mechanically exfoliated on an oxidized silicon substrate. Using a dry transfer technique, we encapsulate the monolayer sample between two hexagonal Boron Nitride (hBN) flakes and place it on a few-layer graphene back gate. The assembled atomic stack is transferred to a silicon chip with two pre-fabricated electrodes for electrostatic gating. Using this procedure we routinely fabricate samples of good quality comparable to other recent works on $MoSe_2$ and other TMDs sandwiched by hBN [31–39]. For optical measurements, the sample is cooled down to 4K and the charge density is tuned by the graphene back gate. Noting that the valence band splitting $\Delta_v$ is large (~200meV, Fig.1a), our hole doping only fills the top valence band.

Figures 2a & 2b show the gate-tuned differential reflectance (DR) and photoluminescence (PL) of the sample. The optical features appear prominently in four energy windows of A:1s (1.6-1.65eV), A:2s (1.75-1.8eV), B:1s (1.8-1.87eV) and B:2s (1.95-2.03eV). Their gate voltage ($V_g$) dependence can be divided into two ranges, $0 < V_g < 2V$ and $V_g < 0$ for before and after the formation of the hole Fermi sea, labeled as 'charge neutral' and 'hole doped' on the heatmaps.

In Fig.2a at charge neutral, four DR resonances are visible: two strong features at ~1.65 and ~1.85eV due to A:1s and B:1s, and two weaker features at ~1.8 and ~2.0eV, attributable to A:2s and B:2s respectively. With hole doping A:1s DR exhibits two repelling branches and B:1s redshifts, while the 2s features of both A and B transitions disappear.

For PL in Fig.2b, three neutral exciton emission features (labeled X) due to A:1s, A:2s and B:1s are observed at positive $V_g$'s (B:2s PL is not visible under this experimental condition). They quench rapidly once $V_g$ becomes negative. For negative $V_g$, we observe two $X^+$ features associated with the A:1s and A:2s states. More detailed PL evolutions are shown in representative spectra in Figs. 2c & 2d. Note that the A:1s emissions are much more intense than those from A:2s and B:1s due to Kasha's rule, similar to our previous study in $WSe_2$ [12]. As a result, A:1s $X^+$ mode emission, although becoming weaker, extends to the nominally charge neutral $V_g > 0$ range.

The A and B transitions involve electronic states with similar wavefunctions except for the spin orientation and minor differences in effective mass. They are expected to have similar Coulomb interaction and binding energies. This agrees with our assignment in DR the 2s states: the B 2s-1s energy separation is similar to that of A 2s-1s, both ~150meV. This assignment is also consistent with several previous studies[33,40,41].

Figure 3 plots the peak energy and linewidth evolution for A:1s and A:2s with doping (filled symbols: PL; open symbols: DR). From the constant X and $X^+$ energy and width in the range $0 < V_g < 2V$, we conjecture that the hole Fermi sea starts to form only when $V_g$ becomes negative. In the nominally charge neutral region, we attribute the A:1s $X^+$ PL emission to A:1s trions, three-particle states formed by the binding of neutral excitons with residual holes that have not yet formed a well-defined Fermi sea. This provides an accurate measurement of A:1s trion binding energy. In Fig.3b we plot the X-X⁺ splitting and find $E_{Tb}^{A:1s}$ = 27meV. The A:2s X⁺ mode is not visible for 0<Vg<2V in either PL or DR. Nevertheless, from PL data at $V_g$ < 0, we can extrapolate from Fig.3b that the A:2s trion binding energy $E_{Tb}^{A:2s}$ is also about 27meV.

The observation of $E_{Tb}^{A:1s} \approx E_{Tb}^{A:2s}$ is somewhat surprising, given that the 2s exciton has a much smaller binding energy than 1s. We performed theoretical calculations for the ground and excited exciton and trion states using the stochastic variational method (SVM) [28,42–44]. The 2D screened electrostatic interaction between the charges is modeled with the Keldysh potential: [45]

$$V(r) = \frac{\pi e^2}{2\epsilon r_0}\left[H_0\left(\frac{r}{r_0}\right) - Y_0\left(\frac{r}{r_0}\right)\right] \quad (1),$$

where $\epsilon = 4.5$ is the dielectric constant of hBN, $r_0$ is the screening length, $H_0$ and $Y_0$ are the Struve and Bessel functions of the second kind, respectively.

Using a reduced exciton mass of $0.35 m_0$, [40] and an electron-hole mass ratio of ~1,[16] we calculate the binding energies and wavefunctions of the 1s and 2s exciton and trion states residing in the hBN sandwiched semiconductor. Treating the screening length as a fitting parameter, we plot in Fig.4a the dependence of calculated 1s and 2s exciton binding energy on $r_0$. Experimentally our 1s and 2s exciton energies are 1643.6meV and 1796.0 meV which gives a 1s-2s splitting of 152.4 meV. Using this splitting, we find $r_0 \approx 1$ nm, and the corresponding 1s and 2s exciton binding energies are 208 and 56 meV respectively.

In Fig.4b, we plot the trion binding energies similarly as a function of $r_0$. The SVM calculations show that there exists one 1s trion and two 2s trion resonances[28]. At small $r_0$, the 1s trion indeed has a larger binding energy than the 2s trion states. However, with increasing $r_0$ the 1s trion binding energy rapidly decreases, and eventually becomes smaller than that of the 2s trions. At ~1nm, the fitted screening length found in Fig.4a from 1s-2s splitting, 2s trion 1 binding energy is about 25meV, while 2s trion 2 and trion 1 are both at ~17meV.

The crossings we observe in Fig.3b suggests that the Coulomb and Keldysh potentials have quite different impacts on the binding of the 3-particle state. Some insights can be gained from the wavefunctions of the exciton and trion states. In Fig.3 panels c-e, we plot the calculated electron-hole and hole-hole correlation functions of the 1s and 2s trions (solid curves), in comparison with the corresponding exciton electron-hole correlation function (dashed curves). The 2s wavefunction is much more extended than 1s. Thus for $r_0 \approx 1$ nm, charges in the 2s trion mostly interact via the 1/r Coulomb tail. For 1s trion however, the correlation functions have a significant weight for $r < 1$nm and interaction strength between electron and holes is highly sensitive to small changes in $r_0$. As $r_0$ varies from 0 to 1 nm, the interaction energies are significantly reduced, leading to rapid decrease of the binding energy, approaching that of the 2s trions. At large $r$, 2s trion 2 has a very long tail between 15 and 35nm not seen in 1s trion and 2s trion 1. This is a very large three-particle state where electron-hole and hole-hole separations are about 5 times of those in 2s trion 1. We are currently not sure whether our 2s X⁺ originates from 2s trion 1 or 2. Mode 1 matches our experimental value better quantitatively, but mode 2 has the same energy as the 1s trion, although the theoretical calculation underestimates the binding strength.

We now examine our data for $V_g < 0$ where a well-defined Fermi sea has formed. As shown in Figs. 2a, 2b, 2c and 3a, the A:1s X and X$^+$ peaks blueshift and redshift respectively in both DR and PL. Quantitatively in Fig.3a, the PL (filled symbols) and DR (open symbols) agree well each with other, suggesting that in contrast to a previous study[22], the emission and absorption originate from the same quasi-particle in the system . Similar repelling behavior is also observed in A:2s; see Figs. 2d and 3a. Note that here we rely mostly on the PL spectra (Figs.2d) since 2s absorption features are weak and difficult to extract the exact mode energy (A:2s X$^+$ is actually not visible).

The energy difference $\Delta E$ between X and X$^+$ modes is plotted in Fig.3b. For A:1s, $\Delta E$ is constant for $V_g > 0$; and for $V_g < 0$, the splitting scales linearly with the Fermi energy:

$$\Delta E_{1s} \approx 27 meV + 0.9 E_F \qquad (2).$$

Similar linear scaling is also seen for A:2s X and X$^+$ mode splitting, albeit with a much larger slope:

$$\Delta E_{2s} \approx 27 meV + 3.9 E_F \qquad (3).$$

It is instructive to compare this $E_F$ dependence with MoS$_2$ where $\Delta E$ for A:1s was found to be $18 meV + E_F$ [15]. Thus the A:1s trion here has a binding energy that is 9meV larger than in MoS$_2$, consistent with previous studies[16,41]. The linear dependences of $\Delta E_{1s}$ on $E_F$ in the two systems agree well with each other. The pre-factor of $E_F$ at a value of ~1 was interpreted as the additional energy required to place the charge from the disassociated trion on the top of Fermi sea due to Pauli blocking.[15] Our observation of ~0.9 in Eqn.(2) is in line with this interpretation. However, this Pauli blocking argument breaks down for A:2s, as one would arrive at an unphysical interpretation that holes disassociated from 2s trions need to be placed 2.9 $E_F$ above the Fermi energy.

Despite the quantitative differences between Eqns. (1) and (2), the qualitative similarity in the doping dependence for A:1s and A:2s suggests that the interaction between excitons and a Fermi sea share the same physics origin. Below, we use the term 'polaron' to denote the dressed quasiparticle in the interacting exciton-Fermi sea system. Theoretically, polaronic dressing of excitons in a Fermi sea of holes was described in Refs. [20,21] by:

$$H = \sum_{\alpha,\mathbf{k}} \epsilon_{\alpha,\mathbf{k}} h^\dagger_{\alpha,\mathbf{k}} h_{\alpha,\mathbf{k}} + \sum_{\alpha,\mathbf{k}} \omega_{\alpha,\mathbf{k}} a^\dagger_{\alpha,\mathbf{k}} a_{\alpha,\mathbf{k}} + \sum_{\alpha,\alpha',\mathbf{k},\mathbf{k}',\mathbf{q}} U_{\mathbf{q},\alpha,\alpha'} h^\dagger_{\alpha,\mathbf{k+q}} a^\dagger_{\alpha',\mathbf{k}'-\mathbf{q}} a_{\alpha',\mathbf{k}'} h_{\alpha,\mathbf{k}} \qquad (4),$$

where $\epsilon_{\alpha,\mathbf{k}}$ and $\omega_{\alpha,\mathbf{k}}$ are the energy of a hole and an exciton of momentum $\mathbf{k}$ at the valley $\alpha$ = +K or -K, $U_{\mathbf{q},\alpha,\alpha'}$ describes the interaction between a hole in valley $\alpha$ and an exciton in valley $\alpha'$ with momentum transfer $\mathbf{q}$. In 1L TMDs where the electron-hole mass ratio does not deviate significantly from 1, the low energy interaction physics can be approximated by dressing excitons with holes in the opposite valley. Further, assuming that the exciton-hole interaction is short range in nature, $U_{\mathbf{q},\alpha,\alpha'}$ is momentum independent and is given by $U_{\alpha \neq \alpha'}$. Using this model, the spectral function of the exciton-polaron system was theoretically calculated [20,21]. In Fig.3a lower subpanel we have reproduced this calculation result as the background grey scale heatmap at $\mathbf{q} = \mathbf{0}$ relevant to our measurements. Note that here we focus on the polaronic interaction effect, and did not include screening of the hole Fermi sea that renormalizes the band gap and exciton binding energy [20]. Other than an overall blueshift attributable to these renormalization effects, the theory captures our A:1s experimental results well.

The linewidth of A:1s X and X$^+$ peaks provide interesting information about the properties of the polarons. At positive V$_g$, X$^+$ is wider than X, reflecting that the trion, as a charged entity, is more sensitive to sample potential fluctuations in the absence of Fermi sea screening. The formation of the hole Fermi sea at $V_g < 0$ has opposite impacts on X and X$^+$ lifetime, broadening the former and narrowing the latter. The narrowing of the X$^+$ peak can be understood as the joint impact of Fermi sea screening and state transition from the trion to the attractive polaron. Theoretical calculations show that the attractive polaron, as the ground state of the exciton-polaron system, should be narrow and the linewidth is $E_F$ independent [20,21]. Our experimental data show that it eventually broadens, indicating presence of additional decay mechanisms not included in the theory. The width of X for $V_g < 0$ increases approximately linearly with the Fermi energy as $\sim 0.7 E_F$, as a result of the decay of the repulsive polarons into attractive polarons while creating Fermi sea fluctuations, anticipated to have a larger phase space as $E_F$ increases. Theoretically the linewidth of the repulsive polaron is expected to increase as $\sim 0.87\ E_F$ [20,21], in reasonable agreement with our experiment.

The Hamiltonian of Eqn.(4) treats the exciton as a rigid impurity regardless of its internal structure, and existing exciton-polaron calculation is for the limit of weakly perturbed exciton. Such calculations require that the exciton binding $E_{Xb}$ to be much larger than $E_{Tb}$ and $E_F$. From our SVM calculations above, the 1s exciton binding energy $E_{Xb}^{A:1s} \approx 208 meV$. This gives $E_{Tb}^{A:1s}/E_{Xb}^{A:1s} \approx 0.13$, and $E_F/E_{Xb}^{A:1s} \approx 0.1$, which are reasonable for the perturbative treatment. The scenario changes for the 2s exciton, for which we find, $E_{Tb}^{A:2s}/E_{Xb}^{A:2s} \approx 0.48$, and $E_F/E_{Xb}^{A:2s} \approx 0.36$. This much larger interaction parameters give rises to more significant polaron dressing effects. Quantitatively for the repulsive and attractive modes, we find $E_r^{A:2s} \approx E_X^{A:2s} + 2.1 E_F$ and $E_a^{A:2s} \approx E_T^{A:2s} - 1.8 E_F$ respectively, with $E_X^{A:2s} = 1.796$ eV and $E_T^{A:2s} = 1.769$ eV. This gives rise to the $3.9 E_F$ dependence in Eqn.(3). The A:2s polaron linewidth broadening scales with the Fermi energy at about 6.6 $E_F$, which is also much more significant than that for A:1s polarons.

We comment that proper theoretical treatment of the 2s exciton-polaron is so far lacking. Summing over more Feynman diagrams of the Hamiltonian in Eqn.(4) could be a first step to count for the larger interaction parameters. Beyond that, one might need to take into consideration that the 2s trion electron-hole correlation function can be appreciably different from that of the 2s exciton, especially for 2s trion 2 shown in Fig. 4e.

It is also of interest to examine B:2s for which the interaction parameter ratios are expected to be similar to those of A:2s. We did not observe though the B:2s PL with 532nm laser excitation in Fig.2b. This is likely because the B:2s exciton energy is above the bare bandgap (~1.85 eV); it can thus decay into free electron-hole pairs and has very short lifetime. We were able to observe B:2s absorption in Fig.2a, but the broadening and the asymmetric lineshape make it difficult to extract the polaron energy reliably. The exciton PL upconversion turns out to be a useful tool here. As shown in Fig.5a, the X branch of B:2s PL becomes visible when we resonantly excited A:1s excitons. This resonant enhancement is similar to previous upconversion of A:2s exciton in 1L-TMDs [33]. Once again, we observe that the peak energy and width are roughly constant when charge neutral ($V_g > 0$), and increase approximately as $2 E_F$ and $12 E_F$ respectively with hole doping, similar to the enhanced exciton-polaron dressing effects observed for the A:2s X branch.

We finally discuss B:1s whose X branch DR displays a puzzling redshift in Fig.1a, at odds with observations for X branches of A:1s, A:2s and B:2s. The PL spectra once again give us more insight: the peaks of B:1s PL in Fig.2d actually blueshift with charge doping, consistent with repulsive polaron behavior. Further, if we inspect the B:1s PL carefully, there is a small shoulder ~27meV below the main peak which could be due to the B:1s trion. With doping however, it becomes difficult to distinguish due to the blue-shifting and more intense A:2s repulsive polaron PL. We speculate that the large linewidth of B:1s polarons masks the splitting between its attractive and repulsive modes, which, combined with the spectral weight transfer, lead to the apparent redshift, mostly due to the attractive polaron.

Taken together, our experimental studies on gated 1L-MoSe$_2$ provide a comprehensive analysis of exciton-polaron effects in the TMD atomic layer. Fundamental features of an exciton interacting with a Fermi sea of charges, such as mode repelling, linewidth broadening, spectral weight transfer, are highly robust. The weak perturbation theory gives good quantitative account of the behavior for the A:1s exciton polarons. However, quantitative understanding of A:2s exciton-polaron, where the interaction parameters are larger, requires further theoretical investigation. Our experimental studies provide a first glimpse into this regime, and our theoretical calculations of the excited trion mode correlation functions indicate that the rigid exciton approximation requires more scrutinization, especially for the 2$^{nd}$ mode where the electron-hole correlation function is drastically different from that of the 2s exciton.

## Acknowledgments

This work is supported by the University of Massachusetts Amherst, and in part by the National Science Foundation (NSF) ECCS-1509599. K.V. was supported by NSF under Grant No. IRES 1826917. K.W. and T.T. acknowledge support from the Elemental Strategy Initiative conducted by the MEXT, Japan, Grant Number JPMXP0112101001, JSPS KAKENHI Grant Numbers JP20H00354 and the CREST(JPMJCR15F3), JST.

## Data Availability Statement

The data that support the findings of this study are available from the corresponding author upon reasonable request.

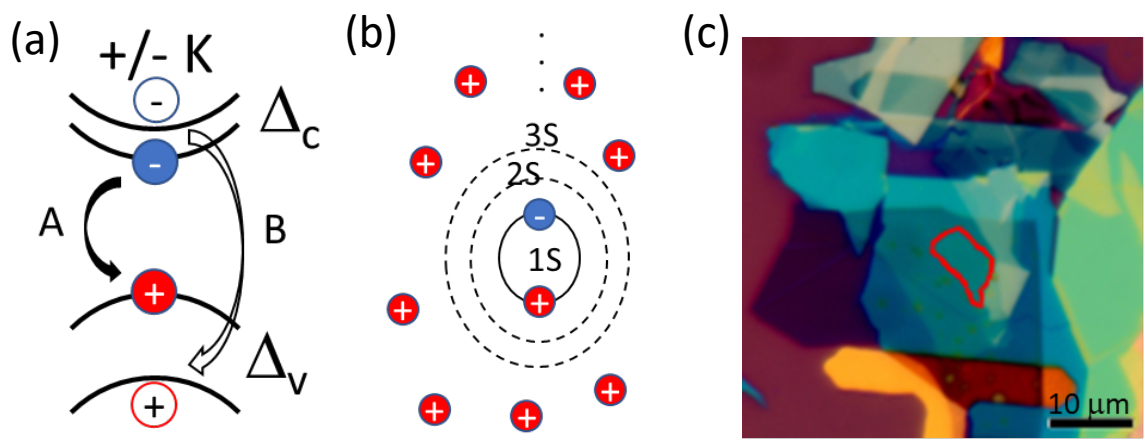

FIG.1 (a) Schematic of TMDC band structure at the K points. Filled/open circles denote charges in A and B excitons. (b) Cartoon of the Rydberg states of an exciton surrounded by a charged sea of holes. (c) Optical microscope image of encapsulated $MoSe_2$ sample with a graphene back gate. The monolayer $MoSe_2$ is highlighted with a red border.

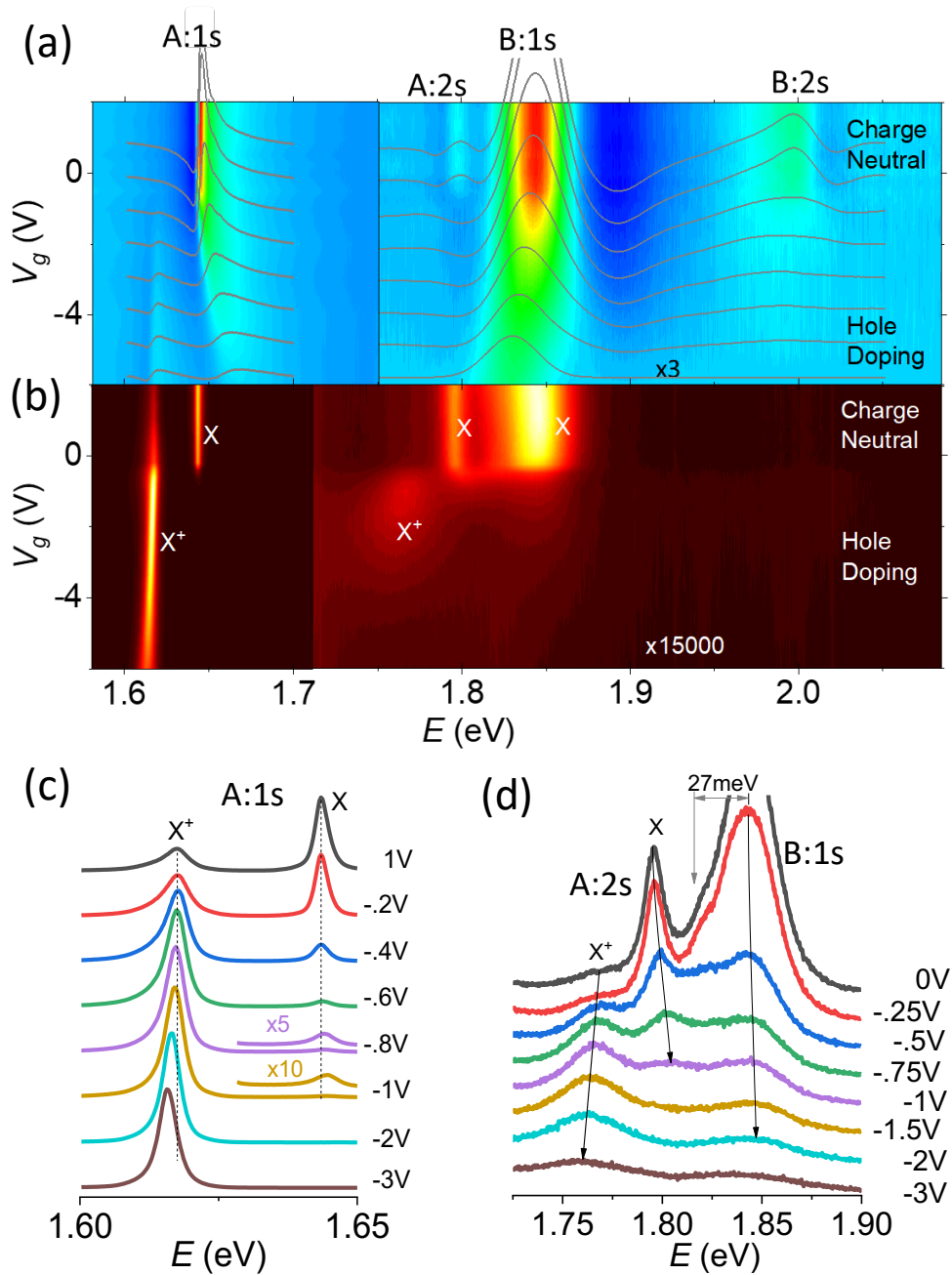

FIG.2 (a,b) Heatmaps of reflection and PL under 532nm excitation, respectively. Negative voltage corresponds to hole doping. (c) Gate dependent A:1S PL spectra. (d) A:2S & B:1S PL spectra, showing redshift of A:2s $X^+$ mode, and blueshift of the A:2S and B:1s X modes. The shoulder 27meV below the B:1s main peak could be due to its trion.

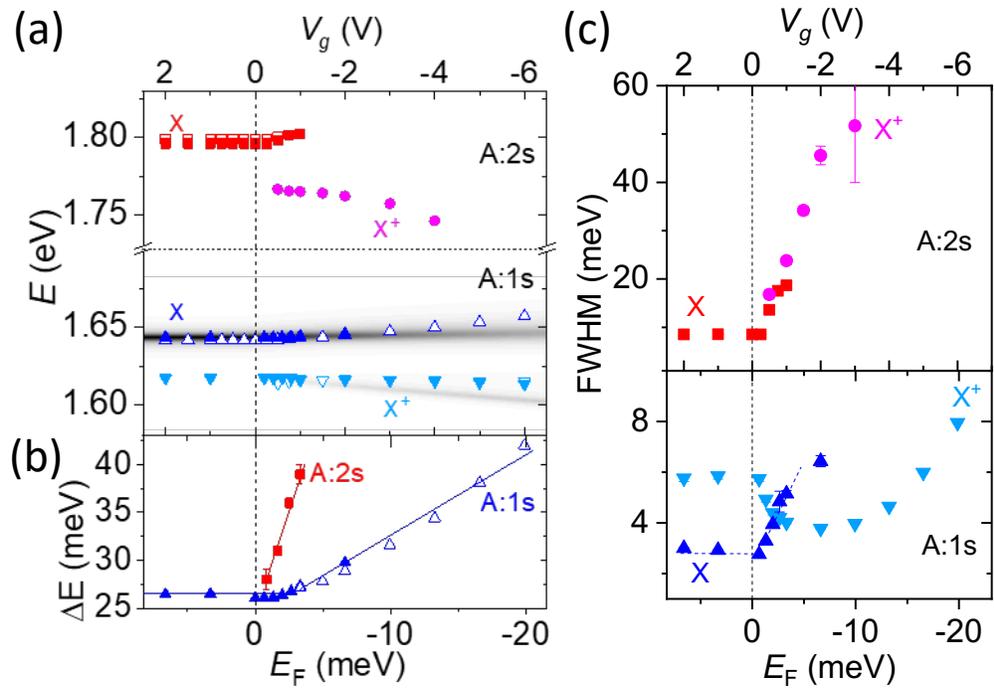

FIG.3 (a) Peak positions as determined by PL (closed symbols) and reflection (open symbols). Superimposed in the lower subpanel is simulated intensity from the exciton-polaron model. Top axis is an estimate of the Fermi energy shift from charge neutrality. (b) The repulsive and attractive polaron splitting as a function of Fermi energy. (c) Doping dependence of A:1s and A:2s PL linewidth.

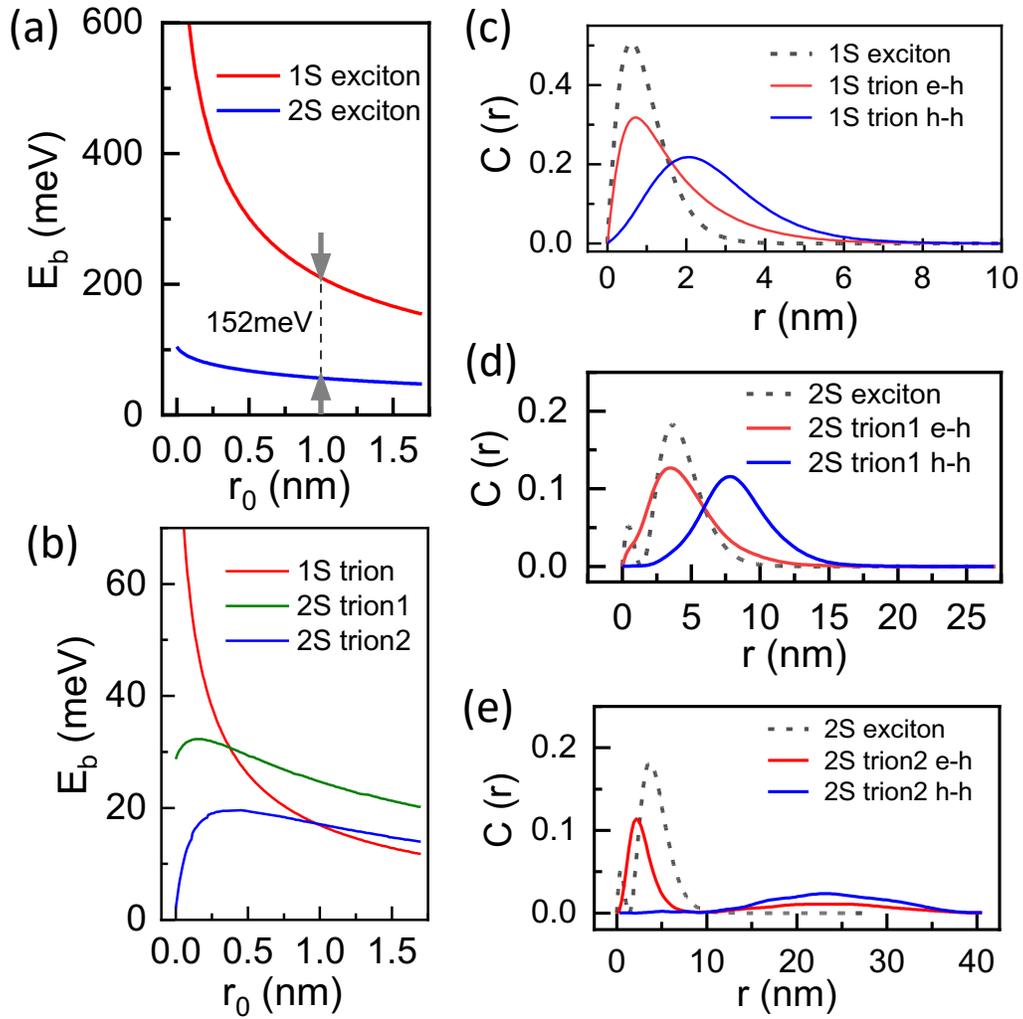

Fig.4 (a) 1s and 2s exciton binding energy as a function of the screening length. Our experimental 152meV 1s-2s separation corresponds to a screening length of ~1nm. (b) Calculated 1s and 2s trion binding energies. Two 2s trion modes are found. (c-e) 1s and 2s trion correlation functions. The dotted curve in the background are 1s and 2s exciton electron-hole correlation function for comparison.

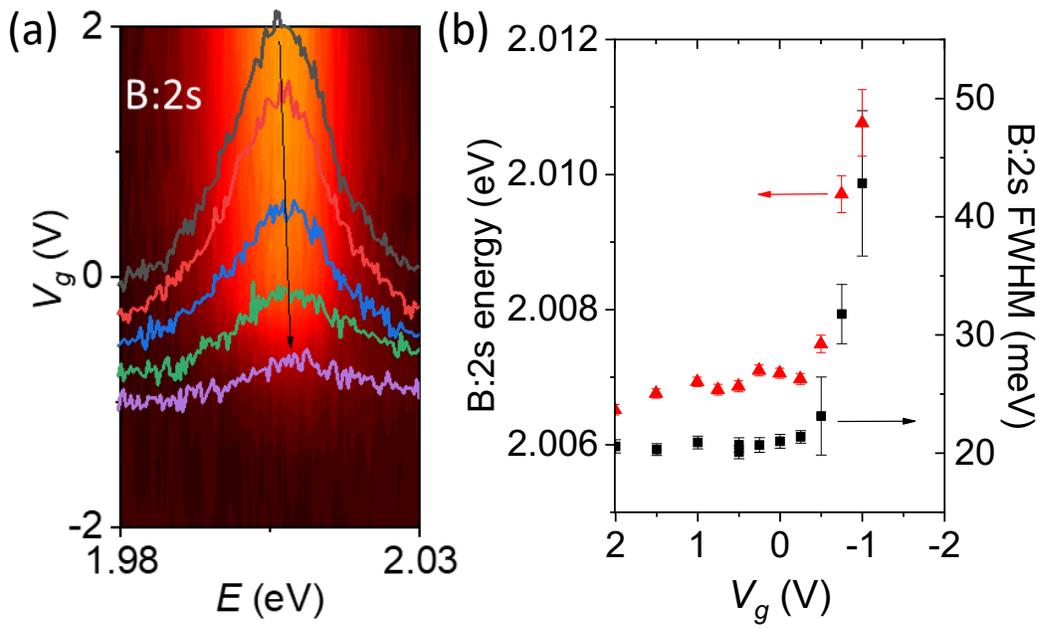

FIG.5 (a) Heatmap of B:2s PL taken by resonant upconversion from populating the A:1s exciton. (b) Fitted width and position of the B:2s on the left and right axes, respectively.